\documentclass[preprint,prb,showpacs,aps]{revtex4}
\usepackage{amssymb}
\usepackage{epsfig}

\begin{document}

\title{Scattering approach to current and noise in interacting mesoscopic systems}

\author{V. Nam Do$^1$, P. Dollfus$^1$, V. Lien Nguyen$^2$}
\affiliation{$^1$ Institut d'Electronique Fondamentale,
B$\hat{a}$timent 220 - UMR8622, CNRS, Universit\'{e} Paris Sud, 91405 Orsay, France\\
$^2$Theoretical Dept., Institute of Physics, VAST, P.O. Box
429  Bo Ho, Hanoi 10000,  Vietnam }

\date{\today}

\begin{abstract}
We propose an extension of the Landauer-Buttiker scattering theory to include effects of interaction in the active region of a mesoscopic conductor structure. The current expression obtained coincides with those derived by different methods. A new general expression for the noise is also established. These expressions are then discussed in the case of strongly sequential tunneling through a double-barrier resonant tunneling structure.
\end{abstract}


\maketitle

\section{Introduction}
Study of electronic transport through a mesoscopic system was
pioneered in  the full quantum mechanical framework by Landauer for a single conduction channel,\cite{Landauer} and then developed by Buttiker for multi-channel and multi-probe systems using the scattering matrix approach.\cite{Buttiker,Martin} The theory, now known as the Landauer-Buttiker scattering theory (or scattering theory, for short), has become one of the standard ways to treat the problem of transport through nano-scale electronic devices, which consist of at least two terminals connected to a domain called scattering or active region. The heart of this theory is the assumption that the scattering region only separates the incident wave into transmitted and reflected parts. Consequently, the structure is characterized by a unique set of transmission $\{T_n\}$ (reflection $\{R_n\}$) coefficients which satisfy the basic probability conservation law, $T_n+R_n=1$. The transport characteristics of the structure are then expressed in terms of these coefficients (see Ref.[\onlinecite{Blanter}] for an overview). The scattering theory has successfully described the steady-state transport through mesoscopic structures in the ballistic regime. However, in most electronic semiconductor devices scattering of carriers on impurities or/and lattice vibrations is usually important, and may even govern the transport process. To solve scattering problems different approaches, such as the non-equilibrium Green's function \cite{Caroli,Lake,Selman,Meir,Ding,Dong} or/and
two-particle Green's function,\cite{Wingreen, Lund} have been developed. Unfortunately, such approaches are often complicated and computationally expensive that it is not easy to gain a quantitative description of interaction effects.

In contrast, the approach based on the scattering matrix is very simple in computation and transparent in physics. Recently, it has been efficiently developed to include the effect of incoherent transport on current-voltage characteristics in nano devices.\cite{Venugopal} To capture such effects, the scattering method has been extended by considering each scattering center as a fictitious probe. The probe is then characterized by a phenomenological parameter interpreted as the local chemical potential, which determines a quasi-equilibrium state of the probe. The conservation of current at each fictitious probe, which is used to determine the local chemical potential, implies that when an electron escapes into the probe, another electron has to be reinjected immediately therefrom to the conductor in such a way that there is no phase correlation between two events.\cite{Buttiker_91} Actually, the scattering region, which includes a lot of scattering centers, can play a role like an especial intermediate source or sink of particles. Similar to the real sources/sinks (the contacts), the current in the intermediate region is conserved globally rather than locally. Moreover, there should be a phase correlation between the incoming and outgoing waves. Therefore, a suitable description is required to feature such an intermediate region. In this work, we develop the scattering method in such a way that it is able to distinguish appropriately the fluxes of coherent and incoherent particles passing through the scattering region. The approach is then applied to discuss the current-voltage ($I-V$) and current noise ($S-V$) characteristics in double-barrier resonant tunneling structures (DBRTS). It results in a current formula coinciding with that available in literature, and a new general formula for the current noise power. From these formulae we can recover all available conclusions of the current and noise from the limit of purely coherent to purely sequential tunneling regime.

The paper is organized as follows: Section II presents our key point for the development of the scattering theory. A current formula is then derived. In Sec. III, we derive a general formula for the shot noise power. We discuss in Sec. IV some important aspects resulting from general expressions obtained in the previous sections. Finally, a brief conclusion is given in the last Sec. V.

\section{Average current formula}
The study is started by considering a structure characterized by a sole lattice temperature, and a chemical potential which can be determined by the applied bias for each real contact. The current, in principle, is the expectation value of the current operator, which is generally written in the form
\begin{eqnarray}
\widehat{I}(z,t) = \frac{\hbar e}{2mi}\sum_\sigma\int
d\textbf{r}_\bot\left[\Psi_\sigma^\dag(\textbf{r},t)
\frac{\partial\Psi_\sigma(\textbf{r},t)}{\partial \textbf{r}}
-\frac{\partial\Psi^\dag_\sigma(\textbf{r},t)}{\partial
\textbf{r}}\Psi_\sigma(\textbf{r},t)\right],\label{1}
\end{eqnarray}
where $\Psi_\sigma(\textbf{r},t)$ is the field operator with the spin-index $\sigma$. Here, we neglect this index for simplicity and replace the sum by the factor $\eta_s=2$ for the spin degeneracy. The transport is assumed to be along one direction so that we can expand the field operators in terms of transverse modes, $\{\chi_{\alpha,n}(\textbf{r}_\bot)\}$, and of incident and outgoing waves,
\begin{eqnarray}
\Psi_\alpha(\textbf{r},t) &=& \int
dEe^{-iEt/\hbar}\sum_{n} \frac{\chi_{\alpha,n}(\textbf{r}_\bot)}{\sqrt{2\pi\hbar
v_{\alpha,n}(E)}}
\left[a_{\alpha,n}(t)e^{ik_{\alpha,n}z}+ b_{\alpha,n}(t)e^{-ik_{\alpha,n}z}\right]\\
\Psi^\dag_\alpha(\textbf{r},t) &=& \int
dEe^{iEt/\hbar}\sum_{n} \frac{\chi^*_{\alpha,n}(\textbf{r}_\bot)}{\sqrt{2\pi\hbar
v_{\alpha,n}(E)}}
\left[a^\dag_{\alpha,n}(t)e^{-ik_{\alpha,n}z}+ b^\dag_{\alpha,n}(t)e^{ik_{\alpha,n}z}\right]
\label{2},
\end{eqnarray}
where the index $\alpha$ is added to refer to the terminal the current is measured in;
$a^\dag_{\alpha,n}(a_{\alpha,n}),b^\dag_{\alpha,n}(b_{\alpha,n})$
are operators creating (annihilating) an electron in an incoming, outgoing state of the mode $n$, respectively; $k_{\alpha,n} = \sqrt{2m(E-E_{\alpha,n})/\hbar^2}$, and $v_{\alpha,n}$ is the velocity. Accordingly, the current operator (\ref{1}) can be rewritten in the form\cite{Blanter}
\begin{eqnarray}
\widehat{I}_\alpha(z,t)&=&\frac{e}{\pi\hbar}\sum_{n}\int dE
dE^\prime e^{i(E-E^\prime)t/\hbar}\nonumber\\
&&\hspace{0.3cm}\left[a^\dag_{\alpha,n}(E)a_{\alpha,n}(E^\prime)
-b^\dag_{\alpha,n}(E)b_{\alpha,n}(E^\prime)\right].\label{3}
\end{eqnarray}

Note that Eq.(\ref{3}) was obtained with the same assumptions  as in the original standard scattering theory. It is now our key point of the analysis. Indeed, in the standard scattering theory one can set a linear relationship between the outgoing wave and the incoming wave amplitudes because of the phase coherence between them. However, in the presence of scattering processes such as the electron-phonon coupling, the phase of electrons is (randomly) changed when the particles jump into the scattering region. When they jump out, they have their own phase which is setup by the interacting processes they have experienced. Accordingly, the scattering region itself should be considered as a quasi-independent intermediate source (sink) providing (absorbing) electrons to (from) the collection (emission) terminal. So that in order to distinguish
the coherent and incoherent particle fluxes through the structure we introduce a couple of operators $d_n$ and $d_n^\dag$ which characterize the processes of annihilating/creating a particle in an interacting state in the scattering region. We thus propose the relation
\begin{eqnarray}
b_{\alpha n} &=&
\sum_{l}\left[\sum_{\gamma}s_{\alpha\gamma,nl}a_{\gamma l}+\lambda_{\alpha,nl}d_{l}\right],\label{4a}\\
b^\dag_{\beta m}
&=&\sum_{k}\left[\sum_{\delta}s^*_{\beta\delta,mk}a^\dag_{\delta
k}+\lambda^*_{\beta,mk}d^\dag_{k}\right]. \label{4b}
\end{eqnarray}
where $[s_{\alpha\gamma,nl}]$ still refers to the scattering matrix and $[\lambda_{\alpha,nl}]$ now refers to the dephasing matrix. The relationship among elements of these matrices is setup by the requirement that the operators (\ref{4a}) and (\ref{4b}) have to satisfy the anti-commutation algebra. That results in
\begin{eqnarray}
&&\sum_{k}\left[\sum_{\gamma}s_{\alpha\gamma,nk}(E)s^*_{\beta\gamma,mk}(E^\prime)
+\lambda_{\alpha,nk}(E)\lambda^*_{\beta,mk}(E^\prime)\right]=\delta_{\alpha\beta}
\delta_{nm}\delta(E-E^\prime).\label{5}
\end{eqnarray}
Obviously, if operators $d^\dag_{m}$ and $d_{m}$ are missing in Eq.(\ref{5})  ($\lambda_\alpha = 0$), the scattering matrix becomes unitary and the usual probability conservation, $T_n+R_n = 1$, is maintained. Under the current consideration, such unitary property is broken, but Eq.(\ref{5}) still leads to a more general conservation law as being seen later. We now denote
\begin{eqnarray}
A^{kl}_{\delta\gamma}(\alpha,E,E^\prime)&=& \delta_{\alpha\delta}\delta_{\alpha\gamma} \delta_{kl}-\sum_{n}
s^*_{\alpha\delta,nk}(E)s_{\alpha\gamma,nl}(E^\prime)\nonumber\\
B^{kl}_{\delta}(\alpha,E,E^\prime)&=&\sum_{n}s^*_{\alpha\delta,nk}(E) \lambda_{\alpha,nl}(E^\prime)\nonumber\\
\overline{B}^{kl}_{\delta}(\alpha,E,E^\prime)&=& \sum_{n}\lambda^*_{\alpha,nk}(E) s_{\alpha\delta,nl}(E^\prime)\nonumber\\
C^{kl}(\alpha,E,E^\prime)&=&\sum_{n}\lambda^*_{\alpha,
kn}(E)\lambda_{\alpha,nl}(E^\prime),\label{6}
\end{eqnarray}
the current operator is then rewritten in the form
\begin{eqnarray}
\widehat{I}_{\alpha}(z,t)&=&\frac{e}{\pi\hbar} \sum_{\delta\gamma}\sum_{kl}\int
dE dE^\prime e^{i(E-E^\prime)t/\hbar}\nonumber\\
&&\hspace{1.5cm}\times
A^{kl}_{\delta\gamma}(\alpha,E,E^\prime)a^\dag_{\delta k}(E)a_{\gamma l}(E^\prime)\nonumber\\
&-&\frac{e}{\pi\hbar}\sum_{\delta}\sum_{kl}\int dE dE^\prime e^{i(E-E^\prime)t/\hbar}\nonumber\\
&&\hspace{1.5cm}\times B^{kl}_{\delta}(\alpha,E,E^\prime)a^\dag_{\delta k}(E)d_{l}(E^\prime)\nonumber\\
&-&\frac{e}{\pi\hbar}\sum_{\delta}\sum_{kl}\int dE dE^\prime e^{i(E-E^\prime)t/\hbar}\nonumber\\
&&\hspace{1.5cm}\times \overline{B}^{kl}_{\delta}(\alpha,E^\prime,E)^*d^\dag_{k}(E)a_{\delta l}(E^\prime)\nonumber\\
&-&\frac{e}{\pi\hbar}\sum_{kl}\int dE dE^\prime e^{i(E-E^\prime)t/\hbar}\nonumber\\
&&\hspace{1.5cm}\times
C^{kl}(\alpha,E,E^\prime)d^\dag_{k}(E)d_{l}(E^\prime).\label{7}
\end{eqnarray}
This equation obviously expresses all possible physical processes of particles such as the transmission from-lead-to-lead, from-lead-to-intermediate region-to-lead, and the relaxation itself. To get the expectation value of this operator it should be noticed that the average is now taken in quantum interacting states due to the presence of incoherent agents such as magnetic impurities, phonons, etc... However, we simply use
\begin{eqnarray}
\langle
a^\dag_{\alpha,n}(E)a_{\beta,m}(E^\prime)\rangle&=&\delta_{\alpha\beta}
\delta_{mn}\delta(E-E^\prime)f_\alpha(E),\\
\langle a^\dag_{\alpha,n}(E)d_{m}(E^\prime)\rangle&=&0,\\
\langle
d^\dag_{n}(E)d_{m}(E^\prime)\rangle &=&\delta_{mn}\delta(E-E^\prime)f_w(E),\label{8}
\end{eqnarray}
where $f_\alpha(E)$ is as usual the Fermi-Dirac function of
electrons  in the $\alpha$-contact, and $f_w$ is referred to the occupation probability in interacting states in the device active region. Note that the appearance of $\delta_{mn}\delta(E-E^\prime)$ in Eq.(\ref{8}) does not mean only elastic processes being included, but it means $f_w$ as the occupation probability of charges in excited states. The average of the second and third terms can be set to 0 because they are equal in magnitude but with opposite sign, and therefore exactly cancel each other. After some simple algebra taking into account Eq.(\ref{5}) the current in the emitter writes
\begin{eqnarray}
I_{e}(t)&=&\frac{2e}{h}\sum_n\int dE
\{[1-R_{e,n}(E)]\left[f_e(E)-f_w(E)\right]\nonumber\\
&&\hspace{2cm}- T_{n}(E)\left[f_c(E)-f_w(E)\right]\}.\label{9}
\end{eqnarray}
Here we have used the eigen-value representation of the matrices $[s^*_{ee}(E) s_{ee}(E)]$ and $[s^*_{ec}(E) s_{ec}(E)]$, which correspond to the reflection ($R_{e,n}$) and transmission ($T_{n})$ probability of a transmission event through the whole structure in the conduction channel $n$. Note that we generally distinguish the reflection coefficients at each contact but, of course, a unique set of transmission coefficients $\{T_n\}$ governs the coherent transport through the whole structure.\cite{Ianna_95}

A similar formula is also easily written down for the current
measured at the collector. The conservation of current allows symmetrizing the current formula by evaluating the occupation factor $f_w$ in the scattering region,
\begin{eqnarray}
f_w =\frac{\Delta_ef_e+\Delta_cf_c}{\Delta_e+\Delta_c},\label{10}
\end{eqnarray}
where $\Delta_{e(c),n}=1-(R_{e(c),n}+T_n) \geq 0$. Substituting this result into Eq.(\ref{9}) yields
\begin{eqnarray}
I_{e}(t)&=&\frac{2e}{h}\sum_n\int dE
\mathcal{T}^{tot}_n(E)\left[f_e(E)-f_c(E)\right],\label{11}
\end{eqnarray}
where
$\mathcal{T}^{tot}_n=T_n+\frac{\Delta_{e,n}\Delta_{c,n}} {\Delta_{e,n}+\Delta_{c,n}}$
regarded as the effective/total transmission coefficient. This result is really interesting because of its simple and well-known form of the Landauer formula. As expected, the total current is a sum of two contributions: the coherent (associated with $T_n$) and sequential (associated with the second term in $\mathcal{T}^{tot}$) parts. Eq.(\ref{11}) leads to the conductance
\begin{eqnarray}
G = \frac {2e^2}{h}\sum_n\left(T_n+
\frac{\Delta_{e,n}\Delta_{c,n}}{\Delta_{e,n}+ \Delta_{c,n}}\right),\label{12}
\end{eqnarray}
which is also coincident with that derived using other methods, such as the voltage-probe method,\cite{Beenakker} and the NEGF method.\cite{Meir}

\section{Current noise power formula}
We now go further to establish a formula for the current noise when both direct and indirect tunneling processes are taken into account. As usual we introduce the operator $\delta
\widehat{I}_\alpha(t)=\widehat{I}_\alpha(t)-\langle
\widehat{I}_\alpha(t)\rangle$, and define the correlation function $S_{\alpha\beta}(t,t^\prime)$ between the currents in contact $\alpha$ and contact $\beta$ as
\begin{eqnarray}
S_{\alpha\beta}(t,t^\prime)=\langle \delta
\widehat{I}_\alpha(t)\delta \widehat{I}_\beta(t^\prime)+\delta
\widehat{I}_\beta(t^\prime)\delta
\widehat{I}_\alpha(t)\rangle.\label{13}
\end{eqnarray}
The current noise power $S_{\alpha\beta}(\omega)$ is then obtained as  the Fourier transform of the time correlation function by noticing that $S_{\alpha\beta}(t,t')$, in fact, depends on $t-t^\prime$.

Using the formula of current operator, Eq.(\ref{7}), calculating the noise  power leads to evaluate the expectation value of the products of two couples of creation and annihilation operators in the interacting states. The products of operators of the same kind can be evaluated using the same algebra as that in the noninteracting case, i.e.,
\begin{eqnarray}
&&\langle
a^\dag_{\alpha,k}(E_1)a_{\beta,l}(E_2)a^\dag_{\gamma,m}(E_3)
a_{\delta,n}(E_4)\rangle \simeq \nonumber\\
&&\delta_{\alpha\delta}\delta_{\beta\gamma}\delta_{kn}\delta_{lm} \delta(E_1-E_4)\delta(E_2-E_3)
f_\alpha(E_1)[1-f_\beta(E_2)] \nonumber\\
&&+\delta_{\alpha\beta}\delta_{\gamma\delta}\delta_{kl}\delta_{mn} \delta(E_1-E_2) \delta(E_3-E_4)f_\alpha(E_1)f_\gamma(E_3), \label{15}
\end{eqnarray}
and
\begin{eqnarray}
&&\langle d^\dag_{k}(E_1)d_{l}(E_2)d^\dag_{m}(E_3)d_{n}(E_4)\rangle
\simeq \nonumber\\
&&\delta_{kn}\delta_{lm}\delta(E_1-E_4)\delta(E_2-E_3)
f_w(E_1)[1-f_w(E_2)]
\nonumber\\
&&+\delta_{kl}\delta_{mn}\delta(E_1-E_2)\delta(E_3-E_4)
f_w(E_1)f_w(E_3),\label{16}
\end{eqnarray}
where $f_w$ is defined by Eq.(\ref{10}). However, it is
inappropriate  to treat similarly the products of two couples of different operators since they manifest the correlation of
charges transferring through different regions of the structure, the many-particle effects. Generally, we set
\begin{eqnarray}
&&\langle a^\dag_{\alpha,k}(E_1)d_l(E_2)d_m^\dag(E_3)
a_{\delta,n}(E_4)\rangle = \delta_{\alpha\delta}\delta_{kn}\delta_{lm}\nonumber\\
&&\times\delta(E_1-E_4)\delta(E_2-E_3)\Xi_\alpha(E_1,E_2),\label{17}
\end{eqnarray}
and
\begin{eqnarray}
&&\langle d_m^\dag(E_1)a_{\alpha,l}(E_2)a^\dag_{\delta,m}(E_3)d_n(E_4)\rangle =
\delta_{\alpha\delta}\delta_{kn}\delta_{lm}\nonumber\\
&&\times\delta(E_1-E_4)\delta(E_2-E_3) \widetilde{\Xi}_\alpha(E_1,E_2),\label{18}
\end{eqnarray}
where $\Lambda_{\alpha} = \Xi_{\alpha}+\widetilde{\Xi}_{\alpha}$ implies the correlation of an occupied state in the contact $\alpha$ with an empty one in the scattering region, and vice versa [see Eq.(\ref{22})]. In the result, we obtain a general formula for the noise power which may  be expressed in four terms
\begin{eqnarray}
&&S_{\alpha\beta}(\omega)=
S^1_{\alpha\beta}+S^2_{\alpha\beta}+S^3_{\alpha\beta}+ S^4_{\alpha\beta},\label{19}
\end{eqnarray}
which are defined as
\begin{eqnarray}
S^1_{\alpha\beta}&=&
\frac{e^2}{\pi\hbar}\sum_{\gamma\delta}\int dE [A_{\gamma\delta}(\alpha,E,E+\hbar\omega)
A_{\delta\gamma}(\beta,E+\hbar\omega,E)]'\nonumber\\
&&\hspace{0.5cm}\times
\{f_\gamma(E)[1-f_\delta(E+\hbar\omega)+ f_\delta(E+\hbar\omega)[1-f_\gamma(E)\},\\
S^2_{\alpha\beta}&=&\frac{e^2}{\pi\hbar}\sum_\gamma\int
dE[B_{\gamma}(\alpha,E,E+\hbar\omega) \times \nonumber\\
&&\hspace{0.5cm}\overline{B}_{\gamma}(\beta,E+\hbar\omega,E)]'
\{\Xi_\gamma(E,E+\hbar\omega)+ \widetilde{\Xi}_\gamma(E+\hbar\omega,E)\},\\
S^3_{\alpha\beta}&=&\frac{e^2}{\pi\hbar}\sum_\gamma\int dE[\overline{B}_{\gamma}
(\alpha,E,E+\hbar\omega)\times \nonumber\\
&&\hspace{0.5cm}B_{\gamma}(\beta,E+\hbar\omega,E)]'
\{\widetilde{\Xi}_\gamma(E+\hbar\omega,E)+ \Xi_\gamma(E,E+\hbar\omega)\},\\
S^4_{\alpha\beta}&=&\frac{e^2}{\pi\hbar}\int dE
[C(\alpha,E,E+\hbar\omega)C(\beta,E+\hbar\omega,E)]'\times \nonumber\\
&&\hspace{0.5cm}\{f_w(E)[1-f_w(E+\hbar\omega)+ f_w(E+\hbar\omega)[1-f_w(E)]\}.
\label{20}
\end{eqnarray}
Here we use the notation $[...]'$ for the trace of matrix in brackets.

As in the standard scattering theory, although $S_{\alpha\beta}$ is written at  arbitrary frequency, equations (\ref{19}-\ref{20}) are only applicable to the case of low frequency (for detailed discussions, see Ref.[\onlinecite{Blanter}]). In this work, we are
interested only in the zero-frequency noise for the two-terminal systems so that we can explicitly achieve a general expression, which reads
\begin{eqnarray}
S_{ee}(0) &&=\frac{4e^2}{h}\sum_{n}\int dE
\left\{T_n[f_e(1-f_e)+f_c(1-f_c)]+ T_n(1-T_n)(f_e-f_c)^2\right\}\nonumber\\
&&+\frac{4e^2}{h}\sum_{n}\int dE
\left\{\Delta^2_{e,n}[f_e(1-f_e)+f_w(1-f_w)]+ \Delta_{e,n}(1-\Delta_{e,n})\Lambda_e\right\}\nonumber\\
&&+\frac{4e^2}{h}\sum_{n}\int dE
\Delta_{e_n}T_n[(f_e-f_c)(1-2f_e)-(\Lambda_e-\Lambda_c)].\label{21}
\end{eqnarray}
It is clear that  each term of Eq.(\ref{21}) can be associated with an explicit and simple physical meaning. Indeed, the first term, which coincides with the main results in
Refs.[\onlinecite{Martin,Buttiker}], and the second one are the thermal and partition noise of the coherent and incoherent tunneling current, respectively, while the third term is interpreted as the correlation between the two current components. Apart from this correlation, i.e., the correlation between the direct and indirect tunneling events, Eq.(\ref{21}) also depends on  $\Lambda_{e(c)}$, the correlation among the charges transferred between the scattering region and the contacts, which is governed by the full statistics of the former region. On average, $\Lambda_{e(c)}$ can be well approximated by
\begin{eqnarray}
\Lambda_{e(c)} = \Lambda_{e(c)}^0 =
f_{e(c)}[1-f_w]+f_w[1-f_{e(c)}],\label{22}
\end{eqnarray}
and our result, Eq.(\ref{21}), then reduces exactly to Eq.(18) in Ref.[\onlinecite{Isawa}], which was derived using the Keldysh's Green functions.

\section{Discussion}
In this section we shall discuss our main results, Eqs.(\ref{11}) and (\ref{21}), derived in the previous sections. It is certainly not necessary to do so for the current formula (\ref{11}) since it absolutely coincides with that obtained by different approaches.\cite{Beenakker,Isawa} So we shall discuss only the noise expression. Although Eq. (\ref{21}) has a simple form and has clear physical meaning for each term, it is difficult to go quantitatively in more detail because of some undetermined phenomenological parameters such as $\Delta_e,\Delta_c$, etc. However, in this section we shall qualitatively discuss some important aspects of this formula. The discussion is for the typical double barrier resonant tunneling structure. After that, we illustrate this discussion using a concrete results obtained in the framework of a microscopic description.

First of all, it is instructive to see a bit the microscopic picture of the electron interaction, e.g., the electron-phonon coupling, in the scattering region. The fact is that an inelastic interaction process may result in significant peaks corresponding to excited states in the density of states (DOS) picture ($E_0+n\hbar\omega_0$, where $\hbar\omega_0$ is the phonon energy, $n$ is an integer number, and $E_0$ is a bound state in the quantum well). The transport of electrons from the emitter to the collector can be then described in the following way. Even the level $E_0$ is not available for the tunneling (lower the emitter conduction band bottom), electrons can jump into the well due to a presence of available excited states in the permitted energy interval. They then either relax to other states in the well before escaping to the collector, or coherently tunnel out from these excited states. Apparently, the height of excited peaks in the DOS determines the coherent tunneling probabilities $\{T_n\}$, i.e., the higher excited peaks, the greater $T_n$ (see Fig. 1). The total current then, as pointed out in the previous sections, can be seen as comprising of two components, sequential and coherent.

Now, look at the third term in Eq. (\ref{21}), it expresses the correlation between the sequential and coherent components of current, and gives rise a negative contribution,
\begin{eqnarray}
S_3 = -\frac{8e^2}{h}\sum_{n}\int dE
\Delta_{e_n}T_n(f_e-f_c)(f_e-f_w).\label{23}
\end{eqnarray}
Here we have used Eq. (\ref{22}). The term becomes important if the coherent tunneling probability $T_n$ is large. Accordingly, if the inelastic interaction process is strong enough, it may cause a reduction of shot noise. While the current increases due to the opening of additional conduction channels, it is reasonable to expect a noise suppression. This point, comment (a), is contradictory to conclusions deduced from the voltage probe approach,\cite{Blanter} but it is commonly accepted (see Ref. \onlinecite{Dong_05}, for instance) and will be illustrated later.

Next, we discuss the case of very small $T_n$, which corresponds to the tunneling induced by very low (insignificant) excited peaks of the DOS. In this limit the first and last terms in Eq. (\ref{21}) can be neglected and we keep only the second one,
\begin{eqnarray}
S_{ee}(0) \approx
\frac{4e^2}{h}\sum_{n}\int dE
\left\{\Delta^2_{e,n}[f_e(1-f_e)+f_w(1-f_w)]+ \Delta_{e,n}(1-\Delta_{e,n})\Lambda_e\right\}.\label{24}
\end{eqnarray}
The quantities $\Delta_e$ and $\Delta_c$ are then directly interpreted as the probability that charges tunnel through the first and second barriers, respectively, from the probability conservation law, $\Delta_{e(c)}+R_{e(c)}\approx1$. Consequently, they can be assumed as constants for simplicity. It should be kept in mind that in the interaction problems effects of many particles generally have to be properly considered. Here, the information on these effects is contained in the corrections to $f_w$ and to $\Lambda_e,\Lambda_c$. Eq. (\ref{24}) can be now estimated for zero temperature and high bias as
\begin{eqnarray}
S_{ee}(0) \propto \frac{4e^2}{h}\frac{\Delta_e\Delta_c}{\Delta_e+\Delta_c} \left[\frac{\Delta_c^2}{\Delta_e+\Delta_c}+ (1-\Delta_e)\frac{\Delta_e+\Delta_c}{\Delta_c}\Lambda_e\right], \label{25}
\end{eqnarray}
and at the same time,
\begin{eqnarray}
I \propto \frac{2e}{h}\frac{\Delta_e\Delta_c}{\Delta_e+\Delta_c}.\label{26}
\end{eqnarray}
The Fano factor therefore is equal to
\begin{eqnarray}
F &=& \frac{\Delta_c^2}{\Delta_e+\Delta_c}+ (1-\Delta_e)\frac{\Delta_e+\Delta_c}{\Delta_c}\Lambda_e\nonumber\\
&=&1-\frac{\Delta_e\Delta_c}{\Delta_e+\Delta_c}+ (1-\Delta_e)\frac{\Delta_e+\Delta_c}{\Delta_c}\delta\Lambda_e.
\label{27}
\end{eqnarray}
Here we have used the expression $\Lambda_e = \Lambda^0_e+\delta\Lambda_e = \Delta_c/(\Delta_e+\Delta_c)+\delta\Lambda_e$. Apparently, if the correction $\delta\Lambda_e$ is neglected, Eq. (\ref{27}) results in a Fano factor smaller than one ($F<1$). However, if the many-particles effects are appropriately taken into account, the following condition can be satisfied:
\begin{eqnarray}
\delta\Lambda_e > \frac{\Delta_e}{1-\Delta_e}\left(\frac{\Delta_c}{\Delta_e+ \Delta_c}\right)^2.\label{28}
\end{eqnarray}
Under this condition, the Fano factor becomes greater than one, $F>1$, i.e., the noise is enhanced. From Eqs. (\ref{27}) and (\ref{28}), we can additionally deduce two following comments: (b) inelastic interaction processes may cause a weak enhancement of the Fano factor, and (c) it is easier to observe $F>1$ in asymmetric structures wherein $\Delta_e > \Delta_c$.

In order to illustrate comments (a), (b) and (c) we use a microscopic description for the process of inelastic electron-phonon interaction developed in Ref. \onlinecite{NAM_07} for the model of a GaAs quantum well embedded between two GaAs/AlGaAs layers. Fig. 1 plots the DOS and the retarded self-energy to show the formation of the excited states due to the interaction. The DOS exhibits a main resonant peak $E_0$ and a significant phonon-emission peak $E_0+\hbar\omega_0$ while the self-energy does the three excited states, one due to the phonon absorption $E_0-\hbar\omega_0$, and two due to the phonon emission (the one with $E_0+\hbar\omega_0$ is strong and the other with $E_0+2\hbar\omega_0$ is weak) which are in accordance to the first and second phonon-assisted tunneling (PAT) regime.

In Fig. 2 we plot two curves of the Fano factor which are extracted for the symmetric case, $\Delta_e=\Delta_c$, and the asymmetric one, $\Delta_e > \Delta_c$. Obviously, the figure shows a suppression of $F$ in the first PAT regime and a weak enhancement in the second PAT one. The figure also shows distinguishably that the noise in the PAT regime is stronger suppressed and higher enhanced in the case of asymmetric compared to the symmetric one, that is consistent with the comment (c).

Thus, we have discussed some aspects of the general noise expression derived in Sec. III. It was shown that the inelastic scattering process may induce either suppression or enhancement of the shot noise (at least in resonant tunneling structures). This conclusion, which was then additionally discussed in the framework of a microscopic description, is not in agreement with those deduced by other approaches in literature. The matter is that in our consideration the role of many-particle effects is strongly emphasized, while in other phenomenological works it is not appropriately taken into account. It is important here to mention that the super-Poissonian noise of $F=1.6$, observed by Li {\it et al.}\cite{LI_90} in GaAs/Al$_{0.4}$Ga$_{0.6}$As-DBRTS, was ensured to come from the electron-phonon interaction process. This experimental data can be seen as a potential support to our discussion on the noise behavior.

\section{Conclusion}
In conclusion, we have developed the scattering theory in an
appropriate  manner such that it can be able to describe the
electron transport through an intermediate region wherein
interacting agents can strongly destroy the phase of incident
electrons or make them relaxing their energy. Our key point here is based on the idea of considering the scattering region as an intermediate source/sink of particles, which maintains the phase correlation among the collision processes. Therefore, we can distinguish the coherent and incoherent particle fluxes which contribute to the total current independently. A derived formula of the current [Eq.(\ref{11})] recovers available results in the literature using different approaches. A new general formula for the current noise power [Eq.(\ref{21})] is also established, from which one can recover well-known results from the limits of purely coherent to purely sequential transport. A short discussion for the case of strongly incoherent tunneling in a double-barrier resonant tunneling structure is also presented and illustrated by a microscopic description.


\newpage
\noindent FIGURE 1: Density of states and retarded self-energy describing the microscopic many-particle picture of the inelastic electron-phonon interaction.

\noindent FIGURE 2: Fano factor curves extracted for the symmetric (Sym.) and asymmetric (Asym.) barriers of the double barrier resonant tunneling structures.

\newpage
\begin{figure}[htp]
\begin{center}
\epsfig{width=8cm,file=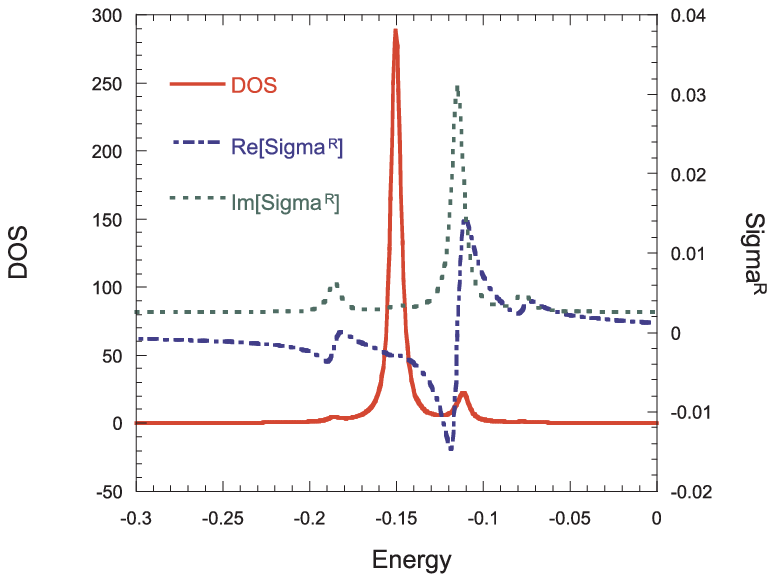,angle=0}%
\end{center}
\end{figure}
\vspace{2cm} FIGURE 1 (V. NAM DO)

\newpage
\begin{figure}[htp]
\begin{center}
\epsfig{width=8cm,file=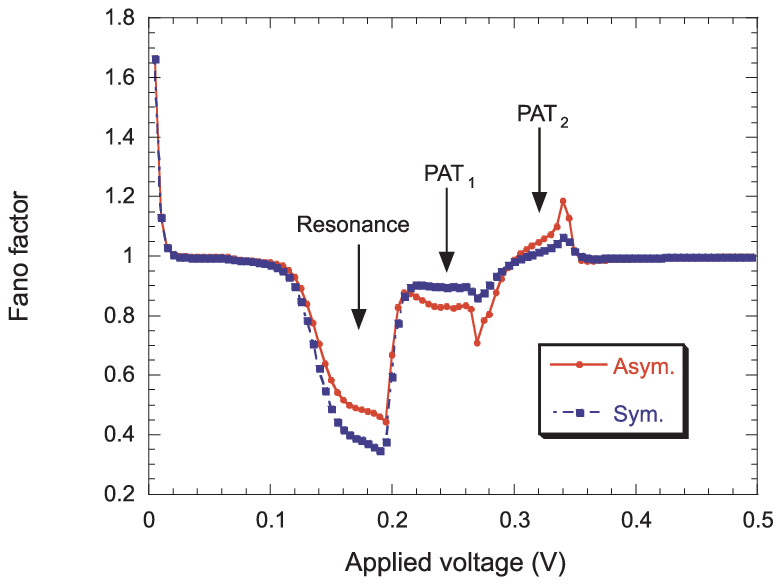,angle=0}%
\end{center}
\end{figure}

\vspace{2cm} FIGURE 2 (V. NAM DO)
\end{document}